\documentclass[aip,pof,longbibliography,twocolumn,reprint]{revtex4-1}

\usepackage{amssymb,amsmath}

\usepackage{epsfig}

\usepackage{graphicx}

\usepackage{color}

\newcommand\beq{\begin{equation}}
\newcommand\eeq{\end{equation}}
\newcommand\beqa{\begin{eqnarray}}
\newcommand\eeqa{\end{eqnarray}}
\newcommand{\al}{\alpha}

\usepackage{epstopdf}

\begin{document}
\title{Comment on ``Kinetic theory models for granular mixtures with unequal granular temperature: Hydrodynamic velocity''
[Phys. Fluids 33, 043321 (2021)]}
\author{Vicente Garz\'o}
\email{vicenteg@unex.es} \homepage{http://www.unex.es/eweb/fisteor/vicente/}
\affiliation{Departamento de F\'{\i}sica and Instituto de Computaci\'on Cient\'{\i}fica Avanzada (ICCAEx), Universidad de Extremadura, E-06071 Badajoz, Spain}

\begin{abstract}

Comment on the paper J. Solsvik and E. Manger, ``Kinetic theory models for granular mixtures with unequal granular temperature: Hydrodynamic velocity,'' Phys. Fluids \textbf{33}, 043321 (2021).

\end{abstract}

\draft

\date{\today}
\maketitle

In a recent work, Solsvik and Manger \cite{SM21} (referred to as the SM-theory) have proposed a kinetic theory for granular mixtures where the velocity distribution functions for each species $f_i(\mathbf{r}, \mathbf{v}; t)$ are assumed to be Maxwellian distributions. Since energy equipartition is broken for inelastic collisions, those distributions are defined in terms of the partial temperatures $T_i$ which are in general different from the granular temperature $T$. In addition, in contrast to the previous works based on the same assumption, the distributions $f_i$ take into account the differences in the mean flow velocities $\mathbf{U}_i$ of the species. Following this approach, the authors evaluate some of the collision integrals appearing in the balance equations for the momentum and kinetic energy. In particular, they obtain corrections to the collision contributions to the momentum and heat fluxes which are of order $|\mathbf{U}_i-\mathbf{U}_j|^2$ and $|\mathbf{U}_i-\mathbf{U}_j|^4$.

On the other hand, a different way of analyzing transport properties in granular mixtures has been carried out in the past few years by Garz\'o, Dufty and Hrenya (the GDH-theory). \cite{GDH07} In contrast to the SM-theory, the GHD-theory solves the Enskog kinetic equation by means of the Chapman--Enskog method up to first order in spatial gradients (Navier--Stokes hydrodynamic order). The transport coefficients are given in terms of the solutions of a set of coupled linear integral equations which are approximately solved by considering the leading terms in a Sonine polynomial expansion. The GDH-theory extends to granular mixtures the results derived years ago for monocomponent granular gases by established kinetic theory models. \cite{GD99} These theories extend to arbitrary inelasticity the results obtained for nearly elastic systems in the seminal works of Jenkins and Savage \cite{JS83} and Lun \emph{et al.} \cite{LSJCh84} The accuracy of the GDH-theory for granular mixtures has been tested with computer simulations and even with real experiments. In the case of computer simulations, the results obtained in the GDH-theory for the tracer diffusion coefficient \cite{GM04} and the shear viscosity coefficient of a heated granular mixture \cite{GM03} show a very good agreement with simulations for conditions of practical interest. In addition, the hydrodynamic profiles derived from the GHD-theory for a single granular fluid compare well with experiments of a three-dimensional system of mustard seeds fluidized by vertical vibrations of the container. \cite{YHCMW02} All these works clearly show the applicability of the GDH-theory for densities outside the low-density regime and values of inelasticity beyond the quasielastic limit.

One of the main deficiencies of the SM-theory is that it does not attempt to solve any kinetic equation since it supposes that the distribution functions of each species are \emph{local} Maxwellian distributions even for inhomogeneous states. In this sense, the SM approach could be considered as a reasonable approach to estimate the collisional transfer contributions to the momentum and heat fluxes but not their kinetic contributions. In particular, the SM-theory yields vanishing Navier--Stokes transport coefficients for \emph{dilute} granular mixtures, which is of course a wrong result.\cite{GD02}

Therefore, it would be convenient to assess the degree of reliability of the SM-theory by comparing its predictions for the collisional contributions to the fluxes with those obtained from the GDH-theory to first-order in spatial gradients. Here, for the sake of concreteness, we will address our attention to the bulk $\eta_b$ and shear viscosity $\eta$ coefficients where computer simulations have clearly shown the accuracy of the GDH-theory, even for strong inelasticity.

According to Eqs.\ (41) and (77) of Ref.\ \onlinecite{SM21}, to first order in gradients, the bulk viscosity $\eta_b$ for a binary mixture of inelastic hard spheres in the SM-theory can be identified as
\beqa
\label{1}
\eta_b^{\text{SM}}&=&\frac{\sqrt{2 \pi}}{9}\sum_{i,j=1}^2 n_i n_j \sigma_{ij}^4\chi_{ij} m_{ij}^2 (1+\al_{ij})\nonumber\\
& & \times \Bigg(\frac{T_i^{(0)}}{m_i}+\frac{T_j^{(0)}}{m_j}\Bigg)^{3/2}\Bigg(\frac{1}{T_i^{(0)}}+\frac{1}{T_j^{(0)}}\Bigg),
\eeqa
where $n_i$ is the number density of species $i$, $\sigma_{ij}=(\sigma_i+\sigma_j)/2$, $\sigma_i$ and $m_i$ being the diameter and mass of particles of species $i$, $m_{ij}=m_i m_j/(m_i+m_j)$, $\chi_{ij}$ is the pair correlation function, $\al_{ij}$ is the coefficient of restitution for collisions $i$-$j$, and $T_i^{(0)}$ is the zeroth-order contribution to the partial temperature of species $i$. In addition, upon obtaining Eq.\ \eqref{1}, use has been made of the fact that the velocity differences $|\mathbf{U}_i-\mathbf{U}_j|$ are at least of first order in spatial gradients so that, $\nabla \mathbf{U}_i=\nabla \mathbf{U}_j=\nabla \mathbf{U}$ and nonlinear terms in $|\mathbf{U}_i-\mathbf{U}_j|$ are neglected in the Navier--Stokes approximation. Here, $\mathbf{U}$ is the mean flow velocity of the mixture. In the SM-theory, the collisional contribution $\eta_c$ to the shear viscosity is simply given by
\beq
\label{2}
\eta_c^{\text{SM}}=\frac{3}{5}\eta_b^{\text{SM}}.
\eeq

The expressions of $\eta_b^{\text{GDH}}$ and $\eta_c^{\text{GDH}}$ in the GDH-theory are more intricate. For  $d$-dimensional granular mixtures, the bulk viscosity $\eta_b^{\text{GDH}}$ can be written as
\beq
\label{3}
\eta_b^{\text{GDH}}=\eta_b'+\eta_b'',
\eeq
where \cite{GDH07,GG19}
\beqa
\label{4}
\eta_b'&=&\frac{\sqrt{2}\pi^{(d-1)/2}}{d^2\Gamma\left(\frac{d}{2}\right)}\sum_{i,j=1}^2 n_i n_j \sigma_{ij}^{d+1}\chi_{ij} m_{ij} (1+\al_{ij})\nonumber\\
& & \times \Bigg(\frac{T_i^{(0)}}{m_i}+\frac{T_j^{(0)}}{m_j}\Bigg)^{1/2},
\eeqa
\beq
\label{5}
\eta_b''=-\frac{\pi^{d/2}}{d\Gamma\left(\frac{d}{2}\right)}\sum_{i,j=1}^2 n_i n_j \sigma_{ij}^{d}\chi_{ij} \mu_{ji} (1+\al_{ij})\varpi_i.
\eeq
Here, $\mu_{ji}=m_j/(m_i+m_j)$ and $\varpi_i$ refers to the first-order contributions to the partial temperatures $T_i$. The quantities $\varpi_i$ have been determined in Ref.\ \onlinecite{GG19} in terms of the parameter space of the mixture. The collision contribution $\eta_c^{\text{GDH}}$ to the shear viscosity is \cite{GDH07,G19}
\beqa
\label{6}
\eta_c^{\text{GDH}}&=&\frac{2\pi^{d/2}}{d(d+2)\Gamma\left(\frac{d}{2}\right)}\sum_{i,j=1}^2 n_i \sigma_{ij}^{d}\chi_{ij} \mu_{ij} (1+\al_{ij})\eta_j^k\nonumber\\
& &+\frac{d}{d+2}\eta_b^{\text{GDH}},
\eeqa
where the kinetic contributions $\eta_i^k$ obey the set of algebraic equations
\beqa
\label{7}
& & \sum_{j=1}^2\left(\tau_{ij}-\frac{1}{2}\zeta^{(0)}\delta_{ij}\right)\eta_j^k=n_iT_i^{(0)}+\sum_{j=1}^2
\frac{\pi^{d/2}}{d(d+2)\Gamma\left(\frac{d}{2}\right)}\nonumber\\
& & \times n_in_j\sigma_{ij}^d m_{ij}\chi_{ij}(1+\al_{ij})\Big[\mu_{ji}\left(3\al_{ij}-1\right)\Bigg(\frac{T_i^{(0)}}{m_i}+\frac{T_j^{(0)}}{m_j}\Bigg)\nonumber\\
& & -
4\frac{T_i^{(0)}-T_j^{(0)}}{m_i+m_j}\Big].
\eeqa
The expressions of the zeroth-order cooling rate $\zeta^{(0)}$ and the collision frequencies $\tau_{ij}$ are given by Eqs.\ (5.51) and (5.65)--(5.66), respectively of Ref.\ \onlinecite{G19}. In addition, the temperature ratio $T_1^{(0)}/T_2^{(0)}$ is determined by equating the partial cooling rates $\zeta_1^{(0)}=\zeta_2^{(0)}=\zeta^{(0)}$. It is important to remark that for elastic collisions ($\al_{ij}=1$) and hard spheres ($d=3$), Eqs.\ \eqref{3}--\eqref{7} of the GDH-theory agree with the results derived many years ago from the Enskog kinetic theory for multicomponent molecular mixtures. \cite{KS79}
\begin{figure}
{\includegraphics[width=0.7\columnwidth]{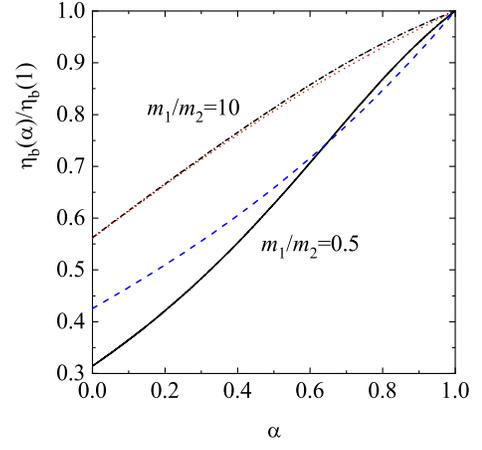}}
\caption{Plot of the (reduced) bulk viscosity $\eta_b(\al)/\eta_b(1)$ versus the common coefficient of restitution $\al$ for $d=3$, $x_1=\frac{1}{2}$, $\sigma_1/\sigma_2=2$, $\phi=0.1$, and two different values of the mass ratio: $m_1/m_2=0.5$ (solid line for the GDH-theory and dashed line for the SM-theory) and $m_1/m_2=10$ (dash-dotted line for the GDH-theory and dotted line for the SM-theory).
\label{fig1}}
\end{figure}
\begin{figure}
{\includegraphics[width=0.7\columnwidth]{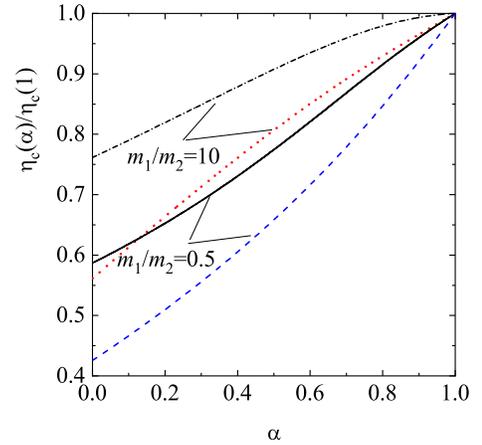}}
\caption{The same as in Fig.\ \ref{fig1} but for the (reduced) collisional shear viscosity $\eta_c(\al)/\eta_c(1)$.
\label{fig2}}
\end{figure}
\begin{figure}
{\includegraphics[width=0.7\columnwidth]{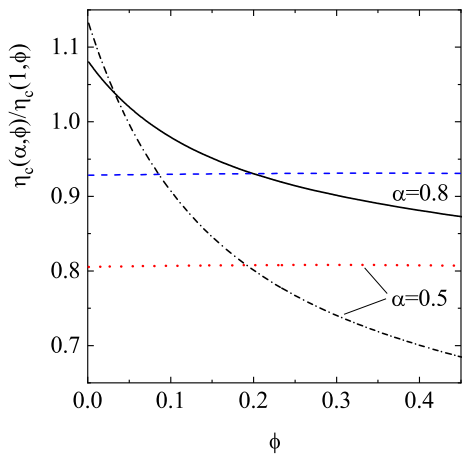}}
\caption{Plot of $\eta_c(\al,\phi)/\eta_c(1,\phi)$ versus the volume fraction $\phi$ for $d=3$, $x_1=\frac{1}{2}$, $\sigma_1/\sigma_2=2$, $m_1/m_2=10$, and two values of $\al$: $\al=0.8$ (solid line for the GDH-theory and dashed line for the SM-theory) and $\al=0.5$ (dash-dotted line for the GDH-theory and dotted line for the SM-theory).
\label{fig3}}
\end{figure}

A comparison between Eqs.\ \eqref{1}--\eqref{2} and \eqref{3}--\eqref{7} show that in general the results obtained for $\eta_b$ and $\eta_c$ from the SM-theory differ from those derived from the GDH-theory, even for elastic collisions. On the other hand, when both the first-order contributions to the partial temperatures $\varpi_i$ and the kinetic contributions $\eta_i^k$ are neglected in Eqs.\ \eqref{3} and \eqref{6}, then the SM and GDH theories agree for elastic collisions for $d=3$. To illustrate the differences between both theories for inelastic collisions, for the sake of simplicity, we consider the case of a common coefficient of restitution ($\al_{11}=\al_{22}=\al_{12}\equiv \al$) of an equimolar mixture ($x_1=n_1/(n_1+n_2)=\frac{1}{2}$) with $\sigma_1/\sigma_2=2$. Moreover, for spheres,
\beq
\label{8}
\chi_{ij}=\frac{1}{1-\phi}+\frac{3}{2}\frac{\phi}{(1-\phi)^2}\frac{\sigma_i\sigma_jM_2}{\sigma_{ij}M_3}+
\frac{1}{2}\frac{\phi^2}{(1-\phi)^3}\Bigg(\frac{\sigma_i\sigma_jM_2}{\sigma_{ij}M_3}\Bigg)^2,
\eeq
where $\phi$ is the total solid volume fraction and $M_s=\sum_i x_i \sigma_i^s$. Figures \ref{fig1} and \ref{fig2} show the $\al$-dependence of the reduced coefficients $\eta_b(\al)/\eta_b(1)$ and $\eta_c(\al)/\eta_c(1)$ for $\phi=0.1$ and two values of the mass ratio. Here, $\eta_b(1)$ and $\eta_c(1)$ refer to the bulk and shear viscosities, respectively, for elastic collisions. It is quite apparent that while the SM-theory reproduces quite well the dependence of $\eta_b$ on $\al$ (specially for $m_1/m_2=10$), important quantitative discrepancies appear in the case of $\eta_c$ for strong dissipation. Much more significant differences between the SM and GDH theories are present in Fig.\ \ref{fig3} where the density dependence of the ratio $\eta_c(\al,\phi)/\eta_c(1,\phi)$ is plotted for two values of $\al$. While the SM-theory predicts a tiny density dependence of the ratio $\eta_c(\al,\phi)/\eta_c(1,\phi)$ for any $\al$, the GDH-theory shows that this ratio decreases with increasing density. The weak influence of the overall solid volume fraction $\phi$ on the (reduced) collisional shear viscosity in the SM-theory is essentially due to the fact that the density dependence of the ratio $\eta_c(\al,\phi)^\text{SM}/\eta_c(1,\phi)^\text{SM}$ is \emph{only} via the partial temperatures $T_i^{(0)}$ (whose dependence on $\phi$ is very small). On the other hand, in the GDH-theory, the dependence of $\eta_c(\al,\phi)^\text{GDH}/\eta_c(1,\phi)^\text{GDH}$ on $\phi$ not only occurs through $T_i^{(0)}$ but also through the kinetic contributions $\eta_i^k$. In particular, in the limiting case of mechanically equivalent particles ($m_1=m_2$, $\sigma_1=\sigma_2$, and $\alpha_{ij}=\alpha$), while the SM-theory predicts that $\eta_c(\al,\phi)^\text{SM}/\eta_c(1,\phi)^\text{SM}=(1+\alpha)/2$ (namely, it is independent of $\phi$), the GDH-theory yields the result
$\eta_c(\al,\phi)^\text{GDH}/\eta_c(1,\phi)^\text{GDH}=A(\alpha,\phi)[(1+\alpha)/2]$, where the function $A$ exhibits in general a complex dependence on both $\alpha$ and $\phi$. In this context, it is worthwhile recalling that the results of the GDH-theory for the shear viscosity $\eta$ agree quite well with computer simulations for moderate densities and/or strong inelasticity (see for instance, Figs.\ 6-8 of Ref.\ \onlinecite{GM03}).

In summary, as expected I have shown that the SM-theory is not able to completely capture the dependence of the bulk and shear viscosities on inelasticity in binary granular mixtures at moderate densities. However, despite these inadequacies, the SM-theory can be still considered as a valuable approach for estimating the collisional contributions to the fluxes since it captures at least qualitatively well (see Figs.\ \ref{fig1} and \ref{fig2}) the $\alpha$-dependence of $\eta_b$ and $\eta_c$.

\textbf{ACKNOWLEDGMENTS}

This work has been supported by the Spanish Government through Grant No. PID2020-112936GB-I00 and by the Junta de Extremadura (Spain), Grants Nos. GR18079 and IB20079, partially financed by ``Fondo Europeo de Desarrollo Regional'' funds.

\textbf{DATA AVAILABILITY}

The data that support the findings of this study are available from the corresponding author upon reasonable request.


\begin{thebibliography}{60}

\bibitem{SM21}J. Solsvik and E. Manger, ``Kinetic theory models for granular mixtures with unequal granular temperature: Hydrodynamic velocity,''
Phys. Fluids \textbf{33}, 043321 (2021).

\bibitem{GDH07}V. Garz\'o, J. W. Dufty, and C. M. Hrenya, ``Enskog theory for polydisperse granular mixtures. I. Navier-Stokes order transport,'' Phys. Rev. E \textbf{76}, 031303 (2007); V. Garz\'o, C. M. Hrenya, and J. W. Dufty, ``Enskog theory for polydisperse granular mixtures. II. Sonine polynomial approximation,'' Phys. Rev. E \textbf{76}, 031304 (2007).



\bibitem{GD99}V. Garz\'o and J. W. Dufty, ``Dense fluid transport for inelastic hard spheres,'' Phys. Rev. E \textbf{59}, 5895 (1999); J. F. Lutsko, ``Transport properties of dense dissipative hard-sphere fluids for arbitrary energy loss models,'' Phys. Rev. E \textbf{72}, 021306 (2005).

\bibitem{JS83}J. T. Jenkins and S. B. Savage, ``A theory for the rapid flow of identical, smooth, nearly elastic, spherical particles,'' J. Fluid Mech. \textbf{130}, 187 (1983).

\bibitem{LSJCh84}C. K. K. Lun, S. B. Savage, D. J. Jeffrey, and N. Chepurniy, ``Kinetic theories for granular flow: inelastic particles in Couette flow and slightly inelastic particles in a general flowfield,'' J. Fluid Mech. \textbf{140}, 223 (1984).


\bibitem{GM04}V. Garz\'o and J. M. Montanero, ``Diffusion of impurities in a granular gas,'' Phys. Rev. E \textbf{69}, 021301 (2004).

\bibitem{GM03}V. Garz\'o and J. M. Montanero, ``Shear viscosity for a moderately dense granular binary mixture,'' Phys. Rev. E \textbf{68}, 041302 (2003).





\bibitem{YHCMW02}X. Yang, C. Huan, D. Candela, R. W. Mair, and R. L. Walsworth, ``Measurements of grain motion in a dense, three-dimensional granular fluid,'' Phys. Rev. Lett. \textbf{88}, 044301 (2002); C. Huan, X. Yang, D. Candela, R. W. Mair, and R. L. Walsworth, ``NMR experiments on a three-dimensional vibrofluidized granular medium,'' Phys. Rev. E \textbf{69}, 041302 (2004).

\bibitem{GD02}V. Garz\'o and J. W. Dufty, ``Hydrodynamics for a granular mixture at low density,'' Phys. Fluids \textbf{14}, 1476 (2002).

\bibitem{GG19}R. G\'omez Gonz\'alez and V. Garz\'o, ``Influence of the first-order contributions to the partial temperatures on transport properties in polydisperse dense granular mixtures,'' Phys. Rev. E \textbf{100}, 032904 (2019).

\bibitem{G19}V. Garz\'o, \emph{Granular Gaseous Flows} (Springer Nature, Cham, 2019).

\bibitem{KS79}J. Karkheck and G. Stell, ``Bulk viscosity of dense simple fluid mixtures,'' J. Chem. Phys. \textbf{71}, 3636 (1979).

\end{thebibliography}
\end{document}